\documentclass[groupedaddress,twocolumn,aps,showpacs,pre,amsmath]{revtex4}
\usepackage{epsfig}
\begin{document}
\title{Self-optimized biological channels in facilitating the transmembrane movement of charged molecules}
\author{V.T.N. Huyen$^1$ , Le Bin Ho$^2$ , Vu Cong Lap$^1$ , and V. Lien Nguyen$^1$ \footnote{ Corresponding author, on leave from VAST/Institute of Physics, Hanoi, email: nvlien@iop.vast.ac.vn}}
\affiliation{$^1$ Institute for Bio-Medical Physics, 109A Pasteur, $1^{st}$ Distr., Hochiminh City, Vietnam \\
$^2$ Hochiminh City Institute of Physics, VAST, Vietnam}
\begin{abstract}
We consider an anisotropically two-dimensional diffusion of a charged molecule (particle) through a large biological channel under an external voltage. The channel is modeled as a cylinder of three structure parameters: radius, length, and surface density of negative charges located at the channel interior-lining. These charges induce inside the channel a potential that plays a key role in controlling the particle current through the channel. It was shown that to facilitate the transmembrane particle movement the channel should be reasonably self-optimized so that its potential coincides with the resonant one, resulting in a large particle current across the channel. Observed facilitation appears to be  an intrinsic property of biological channels, regardless the external voltage or the particle concentration gradient. This facilitation is very selective in the sense that a channel of definite structure parameters can facilitate the transmembrane movement of only particles of proper valence at corresponding temperatures. Calculations also show that the modeled channel is non-Ohmic with the ion conductance which exhibits a resonance at the same channel potential as that identified in the current.
\end{abstract}
\maketitle
\section{Introduction}
Biological channels are responsible for regulating the fluxes of ions and molecules (hereafter referred to as particles for short) across membranes and, therefore, are critically important for the cell functioning \cite{essent}. As well-known, these protein channels are very efficient in the sense that they support a very fast, selective, and robust across-membrane transport, regardless of environment fluctuations \cite{hille}. Surprisingly, such privileged properties have been observed even in the case of large water-filled channels, where the particle transport does not involve the use of metabolic energy or conformational changes and was assumed to be simply diffusive \cite{wickner}. Understanding the nature of this channel-facilitated particle movement (CFPM) is crucially important from the fundamental molecular biology as well as the application point of view (Many modern drugs are developed in the way of using the ion-channels to enhance their efficiency, see for example Refs.\cite{lerche,marban,payan}). 

Experimentally, there are accumulative data showing that the observed CFPM is really resulted from some interaction between the moving particle and the channel-interior lining \cite{meller,moham}. Recent advancements of high-resolution current recording enable single-channel measurements that provide directly a living picture of how an individual channel functions and, therefore, shed light on the characteristics of channel current in dependence on different (channel and environment) parameters \cite{meller,moham,neher}. However, revealing exactly the nature of channel-particle interaction as well as the mechanism of CFPM is still very experimentally problemic due to the puzzled complexities related to both the channel structure and the measurement systems.       

Theoretically, to describe the CFPM several models have been suggested. Considering the one-dimensional (1D) diffusion model with a position-dependent diffusion coefficient, Berezhkovskii et al. supposedly introduced a square potential well, spanning the whole channel length, that brings about a channel-particle interaction  \cite{berezh1,berezh2,berezh3,berezh4}. It was then shown that at a given solute concentration difference there exists an optimum potential well depth that can maximize the particle current, facilitating the channel function. In this model $(i)$ the channel is assumed large enough so that all the effects related to the particle size can be omitted, $(ii)$ a single-particle diffusion is considered, neglecting all many particle correlations, and, particularly, $(iii)$ no realistic potential was assigned as the source for the square potential well introduced. Bauer and Nadler considered a similar 1D diffusion model with a square potential well that is however associated locally with only the particle bound temporarily inside the channel \cite{bauer}. Using the macroscopic version of Fick's equation, it was then demonstrated that a transport increase always occurs for any square potential wells. However, as already noted by the authors, the square potential well exploited in this model is also rather crude and a more realistic potential should be found \cite{bauer}. From the very other point of view, Kolomeisky models the channel as a set of discrete binding sites arranged stochastically \cite{kolomei1}. In such the discrete-state model the particles are assumed to hop along the binding sites in translocations across the channel and the optimum current may be achieved depending on the spatial distribution of binding-sites and the site-particle interactions \cite{kolomei1,kolomei2}. This model is so simple that the main dynamic properties of the problem can be calculated exactly. It was also demonstrated that the discrete-state model \cite{kolomei1} and the continuum diffusion model \cite{berezh1} are closely related and can be effectively mapped into each other \cite{bezruk}. Nevertheless, like the square potential well in the continuum models \cite{berezh1,bauer}, the nature of the binding sites (a kind of channel-particle interaction)  and the hopping mechanism of particles in the discrete-state model \cite{kolomei1} still need to be identified.       

Importantly, in all the models mentioned \cite{berezh1,bauer,kolomei1} the channel-particle interaction (which was expressed by a square potential well or a binding site) is generally viewed as the crucial condition for the transmembrane transport to be facilitated (see also \cite{paglia}). Note again that in these models the particle motion is merely considered one-dimensional. Recently, Dettmer et al. have measured the diffusivity of spherical particles in closely-confining, finite length channels \cite{dettmer}. Measurements demonstrated a strongly anisotropic diffusion in the channel interior: while the diffusion coefficient parallel to the channel axis remained constant throughout the entire channel interior, the perpendicular diffusion coefficient showed an almost linear decrease from the axis towards the channel wall. These observations put forward a need for the two-dimensional (2D) description with direction-dependent diffusion coefficients when studying the movement of particles inside a large channel. Furthermore, experimentally, the single channel kinetics was extensively studied at different external voltages \cite{correa,erdem}. And, the experimental sublinear current-voltage (I-V) characteristics reported in Refs.\cite{anders,busath} is often used as one of the basic requirements for theoretical models \cite{graf}.
 
In the present paper we consider a 2D diffusive movement of particles through a large water-filled channel, taking into account an anisotropy of diffusion coefficients as observed in Ref.\cite{dettmer} and an influence of external voltage as discussed in Refs.\cite{correa,graf}. The channel is modeled as a cylinder characterized by three structure parameters: radius, length, and surface density of negative charges of channel interior-lining. The potential created by this charged interior-lining inside the channel is exactly calculated. It causes the ``channel-particle interaction" that plays a key role in facilitating the transmembrane particle movement. Solving the 2D stochastic Langevin equation for the model suggested we systematically analyze the typically dynamical characteristics of particles such as the translocation probabilities, the translocation times, the currents, and the channel ion conductance under the influence of various factors: the channel-induced potential, the external voltage, or the difference in reservoir particle concentrations. It was particularly shown that to facilitate the transmembrane particle movement the channel should be reasonably self-optimized with appropriate structure parameters so that its potential coincides with the resonant one. In addition, this facilitation is very selective in the sense that a channel of definite structure parameters can facilitate the transmembrane movement of only particles of proper valence at corresponding temperatures. So, the model suggests that facilitating the transmembrane particle movement is an intrinsic property of biological channels. This property is independent of the external factors such as the external voltage or the bulk particle concentration gradient, though these factors may strongly influence the magnitude of various particle dynamical characteristics. 
    
The paper is organized as follows. Sec.II introduces the 2D diffusion model for the problem under study, including the motion equation with an exact expression of the channel-induced potential, and describes the calculating method. Sec.III presents the main numerical results obtained. These results are discussed in great detail, showing the influence of various factors on the particle dynamical characteristics. A particular attention is given to the self-optimized property of the channels in facilitating the transmembrane particle movement. The paper concludes with a brief summary in the last Sec.IV 
\section{Model and calculating method}
We consider a cylindrical channel of length $L$ and radius $R$ that connects the two reservoirs with particle concentrations $n_L$ and $n_R$ as schematically drawn in Fig.1$(a)$. The channel interior-lining carries negative charges which are for simplicity assumed to be continuously and regularly distributed with a surface density $\sigma$. (The cation channels are believed to contain a net negative charge in the pore lining region of the protein \cite{corry}. In the case of potassium and gramicidin channels this is due to the partially charged carbonyl oxygens \cite{essent,corry}). These negative surface charges create an electrostatic potential  $U$ which affects the movement of particles inside the channel. Particles are assumed to diffuse independently, neglecting any many-particle correlation. In addition, the diffusivity of a particle inside the channel is assumed anisotropic with the two different diffusion coefficients,  $D_z$ (parallel with) and $D_x$ (perpendicular to the channel axis). Following  Ref.\cite{dettmer}, we assume that $(i)$ $D_z$ is constant throughout the channel cylinder [$0 \leq |x| < R$ and $0 \leq z \leq L$] and somewhat smaller than the diffusion coefficient $D_0$ in the bulk, $D_z = \alpha D_0$ with $0 < \alpha < 1$ [we choose in the present work $\alpha = 0.5$ for definition] and $(ii)$ $D_x$ linearly decreases as $x$ going from the channel axis (where the diffusion is isotropic) to the channel wall, $D_x = [1 - (| x | / R)] D_z$. 

The model also involves a longitudinal voltage $V$, i.e. the difference in electrical potential between the two channel ends, that may include the intrinsic membrane potential \cite{essent} and/or some externally applied voltage \cite{correa}. This voltage drives the particles moving along the channel. For definition, we assumed that the voltage $V$ is directed from the left to the right [in Fig.1$(a)$] and the charge $q$ carried by a particle is positive. 

Actually, due to the cylindrical symmetry of the channel model suggested, the motion of a particle inside a channel can be effectively described by the 2D stochastic differential equation (Langevin equation for overdamped motion): 

\begin{widetext}
\begin{equation}
\left( \begin{array}{cc}
            \gamma_{zz} &  0   \\
                 0  & \gamma_{xx}
        \end{array}  \right)
\left( \begin{array}{c}
            \dot{z}(t) \\  \dot{x}(t)
        \end{array}  \right) \ = \ -  q \left( \begin{array}{c}
            \partial_z U(x,z)  \\  \partial_x U(x,z)
        \end{array}  \right) + \left( \begin{array}{c}
            q V / L  \\  0
        \end{array}  \right) + \left( \begin{array}{c}
            \xi_z (t) \\  \xi_x (t)
         \end{array}  \right) \ ,
\end{equation}
\end{widetext}

\begin{figure} [t]
\includegraphics{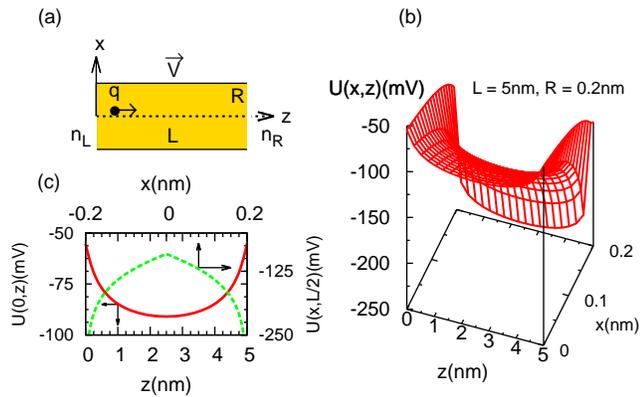}
\caption{
(color online) $(a)$ Model of the cylindrical channel under study; $(b)$ Channel-induced potential $U(x, z)$ of eq.(3) is plotted for the channel with $R = 0.2 \ nm, \ L = 5 \ nm$, and $\sigma = - 0.1 \ C/m^2$ [Note: $U(x,z)$ is symmetrical with respect to the sign of $x$]; $(c)$ the $U(0,z)$-potential well (red-solid line, see the left and bottom axes) and the $U(x,L/2)$-potential barrier (blue-dashed line, see the right and top axes) for the potential $U(x,z)$ in $(b)$. The potential $U_0 \equiv U(0,L/2) \approx -90.6 \ mV$ in this case.}
\label{fig1}
\end{figure}

where $- R < x(t) < R$ and $0 \leq z(t) \leq L$ are the 2D-coordinates of the particle at $t$-time, $\dot{x} \equiv dx / dt$, $\gamma_{xx} \ (\gamma_{zz})$ is the drag coefficient in the $x \ (z)$-direction, $U(x,z)$ is the potential created by the charged channel lining, $q V / L$ is the voltage-induced force acting on a particle of charge $q$ in the $z$-direction, and $\xi_x (t) \ ( \xi_z (t))$ is the random force in the $x \ ( z )$-direction which is as usual assumed to have a zero mean and a white noise correlation:
\begin{equation}
\langle \xi_\nu (t) \rangle \ = \ 0 \ \ {\rm and} \ \ \langle \xi_\nu (t) \xi_\nu (t') \rangle \ = \ 2 D_\nu \delta (t - t') , \ \  \nu = x, z .
\end{equation}
It is here worthy to mention the Stokes-Einstein relation between the diffusion coefficient $D$, the drag coefficient $\gamma$, and the absolute temperature $T$ of a medium: $D \gamma = k_B T $, where $k_B$ is the Boltzmann constant.

In eq.(1) we need to identify the potential $U(x,z)$ inside the channel. Within the model considered, as mentioned above, $U$ is the electrostatic potential created by the charged lining of a cylindrical channel. By solving the fundamental electrostatic problem for a charged cylinder of finite sizes, we can exactly derive an analytical expression of $U$ as a function the $(x, z)$-coordinates [$0 \leq x < R$ and $0 < z < L$]:
\begin{widetext}
\begin{eqnarray}
U(x,z) \ = \  \frac{R \sigma}{\pi \epsilon_0 \epsilon}  \{  ( 1&+&\frac{\pi}{2} )  \ln [ \frac{ z + \sqrt{(x - R)^2 + z^2}}{z - L + \sqrt{(x - R)^2 + (z - L)^2}} ]     \nonumber  \\
 & - & \ln [ \frac{ z + \sqrt{(x + R)^2 + z^2}}{z - L + \sqrt{(x + R)^2 + (z - L)^2}} ]  \  \} , 
\end{eqnarray}
\end{widetext}
where $R, \ L$, and $\sigma$ are the channel structure parameters defined above, $\epsilon_0$ is the vacuum permittivity, and $\epsilon$ is the dielectric constant of the water in the interior of the channel \cite{note02}. Note that the potential $U(x,z)$ is symmetrical with respect to the sign of $x$.

As an example, Fig.1$(b)$ shows the potential $U(x,z)$ of eq.(3) for the channel with $R = 0.2 \ nm, \ L = 5  \ nm$, and $\sigma = - 0.1 \ C/m^2$. At a given $x$-coordinate, $U(z)$ behaves as a symmetrical potential well with the absolute minimum at $z = L / 2$. On the contrary, given a $z$-coordinate, the $U(x)$-curve describes a symmetrical potential barrier with the absolute maximum at $x = 0$ [see, for example, $U(0,z)$ as a function of $z$ (bottom and left axes) and $U(x,L/2)$ as a function of $x$ (top and right axes) in Fig.1$(c)$]. While the well shape of the channel potential $U(x,z)$ in the $z$-direction directly affects the movement of particles across the channel (as will be seen below), its barrier shape in the $x$-direction demonstrates a noticeable role of the transverse motion in the anisotropic 2D-diffusion model considered.  

As a consequence of the observed symmetrical shape, the potential $U(x,z)$ can be characterized by its value at the center of the channel, $(x = 0, z = L/2 )$, where
\begin{equation}
 U(0, L/2)  \ =  \ \frac{R \sigma}{2 \epsilon_0 \epsilon } \ln [ 
     \frac{\sqrt{4 R^2 + L^2} + L}{\sqrt{4 R^2 + L^2} - L} ]  \ \equiv \ U_0 .
\end{equation}
This potential value $U_0$ is uniquely determined by the channel structure parameters ($L, \ R$, and $\sigma$) and can be used to characterize the potential $U(x,z)$ on the whole: each channel creates a unique $U(x,z)$ and each $U(x,z)$ has a unique $U_0$. As an intrinsic characteristics of the channel, the quantity $U_0$ will be used below as a typical measure of the channel potential $U(x,z)$. Fig.1$(c)$ indicates the potential $U_0 \approx - 90.6 \ mV$ for the channel potential $U(x,z)$ examined in this figure. 

Thus, as an extension of the model suggested by Berezhkovskii et al.\cite{berezh1,berezh2,berezh3,berezh4}, the present model is distinguished by the main factors as follows: $(i)$ the diffusion is anisotropically two-dimensional (see eq.(1)), $(ii)$ the negatively charged channel interior-lining creates inside the channel a potential that leads to the first term in the right hand of eq.(1) and that can be exactly identified as a function of only channel structure parameters (see eq.(3)), and $(iii)$ the external voltage causes a driving force expressed by the second term in the right hand of eq.(1). Further, the study will be focused on showing how these factors affect the dynamical characteristics of particles moving through the channel. The dynamical characteristics we are here interested in include the translocation probabilities, the translocation times, the particle current, and the ion conductance. To calculate these quantities we have to solve eq.(1). Reasonably, this stochastic equation can be solved numerically by using the molecular dynamics method \cite{moldyn}. 

A particle enters the channel from either the left ($z = 0$) or the right ($z = L$) at random with the probabilities proportional to the reservoir particle concentration $n_L$ or $n_R$, respectively [Fig.1$(a)$]. The initial $x$-coordinate ($- R < x(t = 0) < R$) and the initial velocity components ($\dot{z}(0)$ and $\dot{x}(0)$) are randomly given, following the standard molecular dynamics simulation procedure \cite{moldyn}. Started from the given initial conditions, a discrete trajectory of the particle is step by step constructed. Given the channel potential  $U(x,z)$ and the external voltage $V$, in each time-step ($\Delta t$) the random forces, $\xi_{x(z)}$, are independently generated and then the final coordinates and velocity of the particle are determined from eq.(1) using the well-known Euler scheme \cite{euler,fathi}. In the $x$-direction the full reflection condition is applied every time when a particle runs into the channel wall, $x = \pm R$. In the other direction, once the $z$-coordinate is out of the range $[0, L]$, the data for the simulated particle is fixed and this particle is no longer followed. The next particle enters the channel and undergoes a diffusion process in the same way as described above. The number of particles involved in getting each of average values of studied dynamical quantities is so large that for all the data points presented below the error bar nowhere exceeds the symbol size [$\approx 10^5$ to $10^7$ particles depending on the quantity and the direction of movement investigated]. The time step is taken to be $\Delta t = 0.0005 \tau_0$, which is believed small enough. The dynamical quantities we are interested in, as mentioned above, include the translocation probabilities, the average translocation times, the net particle current, and the ion conductance.

Actually, the calculating method we exploit in this study is the Brownian dynamics. By solving the Langevin equation, this method is rather appropriate for the problem of interest. A systematical classification of computational approaches proposed and employed for studies of ion channels can be found in  the review paper \cite{maffeo}. Here, in solving numerically eq.(1), for convenience we  choose $L$ as the unit of length, $\tau_0 = L^2 / D_0$ as the unit of time, and $k_B T $ as the unit of energy. So, for example, if $L = 5 \ nm$ and $D_0 = 3.10^{-10} m^2 /s$ \cite{bezruk}, then $\tau_0 \approx 8.3 \ 10^{-8} s$. Remind that $k_B T \approx 8.617 \ 10^{-5} \ eV$ for $T = 1^\circ K$. 
\section{Numerical results and discussions}
In presenting simulation results we introduce for short the symbols $u_0 \equiv qU_0 / k_B T$ (referred to as the effective channel  potential) and $v \equiv q V / k_B T$ (referred to as the effective external voltage). So, the defined parameters $u_0$ and $v$ also contain the particle charge $q$ and the medium temperature $T$. We should keep this in mind when discussing the role of the channel potential in facilitating the transmembrane particle movement. Additionally, for definition, in all the figures relating to the translocation probabilities and the average translocation times the parameters $R$ and $n_{L(R)}$ are kept constant: $R = 0.04 $ (in unit of $L$) and $n_{L(R)} = 145 (15) \ mM $ \cite{essent}. Influences of these parameters will be later discussed when analyzing the net current [Fig.6]. 

Let us first examine obtained results for the translocation probabilities which are separately calculated for the particles moving through the channel from the left to the right ($P_L$) and for those moving in the opposite direction ($P_R$) [see Fig.1 with the $V$-direction indicated]. In simulations, the probability $P_L$ (or $P_R$) is determined as the ratio of the number of particles that passed through the channel to the total number of particles that entered the channel from the left (or right). 

\begin{figure} [t]
\includegraphics{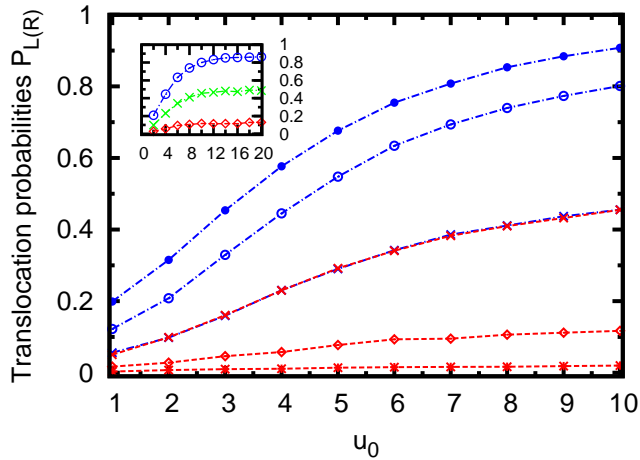}
\caption{
(color online) Translocation probabilities $P_L$ (blue dash-dotted lines) and $P_R$ (red dashed lines) are plotted versus $u_0 \equiv qU_0 /k_B T$ for the channels of the same $R, \ n_L$, and $n_R$, but at different voltages $v \equiv q V / k_B T$:  $0 (\times ), \ 2 (\circ$ and $\diamond )$, and $4 (\bullet$ and $ \ast )$. The points are the simulation results, whereas the lines are drawn as a guide for the eyes [$ R = 0.2 \ nm, \ L = 5 \ nm,  \ n_L  = 145 \ mM, / n _R = 15 \ mM$]. }
\label{fig2}
\end{figure}

Fig.2 shows $P_L$ (blue dash-dotted lines) and $P_R$ (red dashed lines) plotted versus $u_0$ for the channels at different effective voltages $v$:  $0 (\times ), \ 2 (\circ$ and $\diamond )$, and $4 (\bullet$ and $ \ast )$. Generally, this figure demonstrates that with increasing $u_0$ both the translocation probabilities, $P_L$ and $P_R$, increase steadily first [see main figure] and then become saturated [see the inset]. Such the $P_{L(R)}$-versus-$u_0$ behavior is observed at any voltage $v$. In the case of zero $v$, due to the left-right symmetry of the potential $U(x,z)$ of eq.(3) the two curves, $P_L$ and $P_R$, are totally coincidental and the common curve may be in a qualitative comparison with Fig.3 in Ref.\cite{berezh1}(where the considered diffusion is one-dimensional and the potential well is square). Note that with the chosen direction of $V$ [Fig.1, $q$ is positive] the external voltage raises $P_L$ (two higher curves) while suppressing $P_R$ (two lower curves), compared to the case of $v = 0$ (the middle curve).   

\begin{figure} [t]
\includegraphics{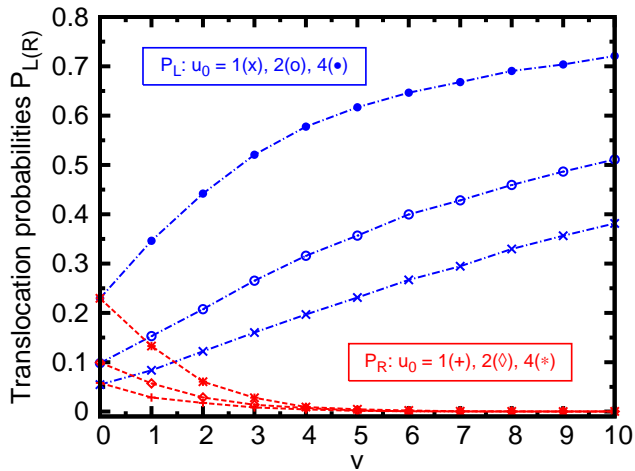}
\caption{
(color online) Translocation probabilities $P_{L}$ (blue dash-dotted lines) and $P_{R}$ (red dashed lines) are plotted versus $v$ for the channels with different effective potentials  $u_0$:  $1 (\times$ and $+ ), \ 2 (\circ$ and $\diamond )$, and $4 (\bullet$ and $ \ast )$. Other parameters and symbols are the same as in Fig.2. }
\label{fig3}
\end{figure}

The external voltage effects can more clearly be seen in Fig.3 where the probabilities $P_{L(R)}$ are presented as the functions of $v$ for the channels with different $u_0$: $1 (\times $ and $+$); $2 (\circ$ and $\diamond$); and $4 (\bullet$ and $\ast$). At zero $v$ the two probabilities $P_{L(R)}$ associated to the same $u_0$ are of equal value [two corresponding curves are started from the same point]. With increasing $v$ the probability $P_L$ smoothly rises, while the probability $P_R$ strongly descends. At $v \geq 5$ the probabilities $P_R$ become practically vanished for all the channels under study [no particle can move through the channel in the right-to-left direction]. The probability $P_L$, on the contrary, continues to grow with the tempo that gradually slows down at higher $v$. Calculations reveal that even at $v = 100$ the channels are still not perfectly transparent for the positively charged particles moving along the external voltage direction [$P_L = 0.98$ or 0.95 for $u_0 = 4$  or 1, respectively].

Next, we consider another fundamental characteristics - the average translocation time. In accordance with the probabilities $P_{L(R)}$ studied in Figs.2-3, we separately calculated the average translocation times for the particles moving through the channel from the left to the right ($\tau_L$) and for those moving in the opposite direction ($\tau_R$). In simulations, we count the time each of simulated particles spends inside the channel. The average translocation time $\tau_L$ (or $\tau_R$) is then obtained by averaging these spending times over all the particles that passed through the channel from the left to the right (or from the right to the left). 

\begin{figure} [t]
\includegraphics{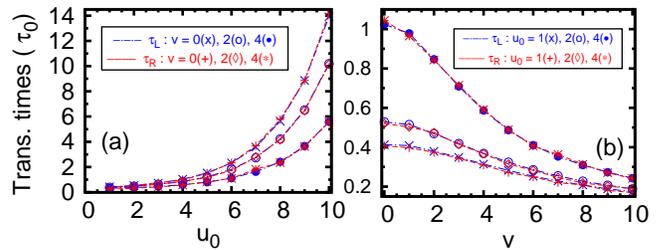}
\caption{
(color online) Translocation times $\tau_L$ (blue dash-dotted lines) and $\tau_R$ (red dashed lines) are plotted versus $u_0$ at different $v$:  $0 (\times$ and $+ ), \ 2 (\circ$ and $\diamond )$, and $4 (\bullet$ and $ \ast )$ $(a)$ and $\tau_{L(R)}$ versus $v$ at different $u_0$:  $1 (\times$ and $+ ), \ 2 (\circ$ and $\diamond )$, and $4 (\bullet$ and $ \ast )$] $(b)$. Other parameters are the same as in Fig.2. }
\label{fig4}
\end{figure}

Fig.4 shows how obtained translocation times $\tau_{L(R)}$ vary with the effective potential $u_0$ [Fig.4$(a)$] or the effective voltage $v$ [Fig.4$(b)$]. Interestingly, in all the cases studied in both the figures, Fig.4$(a)$ and Fig.4$(b)$, the two points, corresponding to $\tau_L$ and $\tau_R$, are practically coincided. So, our 2D-simulations suggest a general equality, $\tau_L (u_0 , v) = \tau_R (u_0 , v)$, that should be always valid in the model studied regardless of the shape of the channel potential $U(x,z)$ as well as the presence of the external voltage $V$. This really causes some surprise, noting on the directed influence of the voltage $V$. Actually, a similar equality of the two average translocation times has been previously suggested in Ref.\cite{berezh2}, but it was there relating to the 1D diffusion model without any external voltage. Fig.4 thus allows us to deal with  the two times $\tau_L$ and $\tau_R$ as a single average translocation time that will be below denoted simply by $\tau$.

The fact that the channel potential $u_0$ raises the translocation time $\tau$ in Fig.4$(a)$, while it also raises the translocation probabilities in Fig.2, might cause some surprise. Actually, as will be seen below, it turns out that a competition between these two seemingly contrary effects of the potential   $u_0$ leads to the most important phenomenon in the ion-channel physics - the CFPM. Comparing the points from three curves with different voltages $v$, we learn that in the region of large $u_0 \ [u_0 \geq 6$ in Fig.4$(a)$] the time $\tau$ decreases almost linearly as $v$ increases from $0$ to $4$. In a wider range of $v$, Fig.4$(b)$ shows that the larger the effective potential $u_0$, the stronger the relative effect of $v$ on $\tau$ becomes. In the limit of high external voltage when the $v$-induced driving force becomes to dominate the right hand in eq.(1), the translocation time should depend on $v$  as $\tau \propto 1 / \sqrt{v}$.

While the question of the particular time that most relevantly describes the transmembrane transport and that can be directly measured is still under discussion \cite{bauer}, the net current has always served as the most important quantity that should be determined theoretically in close comparison with experimental measurements. For the problem under study, the net particle current is determined as the average number of particles the two reservoirs actually exchanged via the channel in a unit of time ($\tau_0$).
 
\begin{figure} [t]
\includegraphics{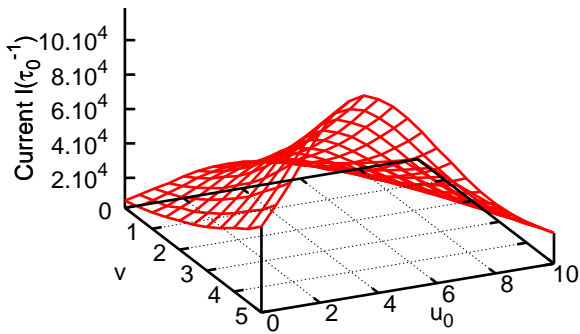}
\caption{
 (color online) 3D-plot of the current $I$ as a function of $u_0$ and $v$. Note on the resonant behavior of the $I$ versus $u_0$ curves at different voltages $v$. Other parameters are the same as in Fig.2 }
\label{fig5}
\end{figure} 
      
Fig.5 presents a 3D-plot of the current $I$ in dependence on the effective channel potential $u_0$ and the effective external voltage $v$. Remarkably, contrary to the monotonic behaviors of $P_{L(R)}$ and $\tau$ in Figs.2-4, Fig.5 shows clearly a resonant behavior of the current $I$: for a given voltage $v$ in the $I$ versus $ u_0$ curve there always has an impressively absolute maximum at some resonant channel potential, $u_0 = u_m$. Remind that $u_0 \equiv q U_0 / k_B T$ with $U_0$ uniquely determined by the channel structure parameters ($L, \ R$, and $\sigma$). So, the maximum observed in Fig.5 implies that for given $q$ and $T$, to successfully facilitate the transmembrane particle movement the channel has to be optimized with the appropriate structure parameters so that its potential $U(x,z)$ coincides with the resonant one. For example, for $q = 1$ and $T = 300 K$, to own the resonant potential of $u_m = 3.5$ as seen in Fig.5, the channel should be self-optimized with the following structure parameters: $L = 5 \ nm, \ R = 0.2 \ nm$, and $\sigma = - 0.1 \ C/m^2$ (given $\epsilon = 80$ \cite{note02}). 

\begin{figure} [t]
\includegraphics{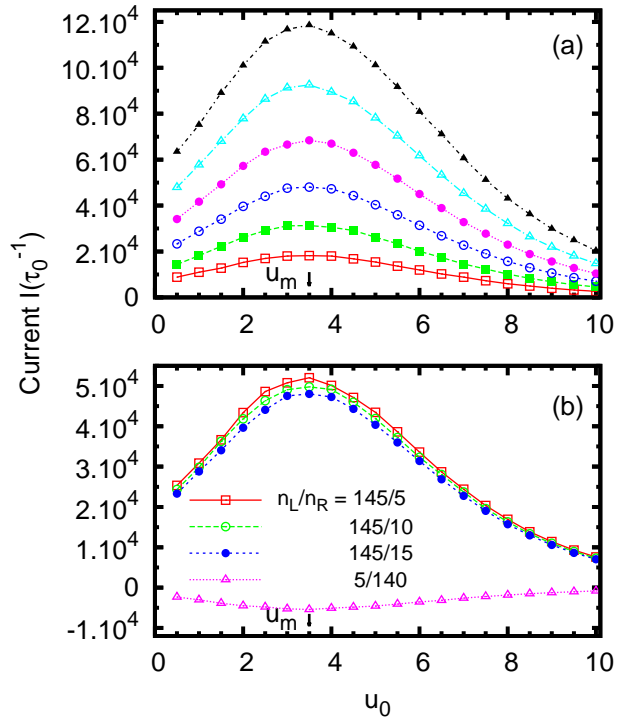}
\caption{
(color online) Resonant channel potential $u_0 = u_m$ is an intrinsic characteristics of the channel. $(a)$ $I$ versus $u_0$ curves extracted from Fig.5 for some values of $v$ [from bottom: $v = 0, \ 1, \ 2, \ 3, \ 4$, and 5]; All these curves show their maximum at the same resonant potential, $u_0 = u_m$ (indicated by the arrow). $(b)$ $I$ as a function of $u_0$ at $v = 2$ for various values of the ratio $n_L / n_R$ [from top: $n_L / n_R= 145/5, \ 145/10, \ 145/15$, and $5/140$]; The resonant potential $u_0 = u_m$ (indicated by the arrow) is independent of reservoirs particle concentration ratio and coincides with $u_m$ determined in Fig.6$(a)$. Note: in the case of $n_L / n_R = 5/140$ the current is negative (flowing from right to left in Fig.1$(a)$) and reaches the largest magnitude at the same $u_0 = u_m$. }
\label{fig6}
\end{figure} 

To see whether the resonant potential $u_m$ depends on the external voltage $v$, we depict in Fig.6$(a)$ some $I(u_0 )$-curves extracted from Fig.5 at various $v$. Surprisingly, the resonant channel potential $u_m$ (indicated by the arrow) is practically the same for all the curves at different voltages $v$. Actually, the fact that $u_m$ is independent of $v$ can be seen right in Fig.5 for  all the values of $v$ under study. Further,  we check if the resonant potential $u_m$ depends on another important external parameter, the difference in particle concentration between the two reservoirs. Fig.6$(b)$ presents the $I(u_0 )$-dependence for several values of the ratio $n_L / n_R$. In the cases of $n_L / n_R = 145/5, \ 145/10$, and $145/15$ (taken from Table 12.1 in Ref.\cite{essent}), all the particle concentration gradients are directed along the external voltage $V$, i.e. from the left to the right in Fig.1$(a)$, and, therefore, the currents are always positive [see the higher three curves]. On the contrary, in the case of the lowest curve in Fig.6$(b)$ for $n_L / n_R = 5/140$ (e.g. for $K^+$-channels \cite{essent}), the particle concentration gradient is directed from the right to the left in Fig.1$(a)$ and, consequently, the current becomes negative (Note that in this case the concentration gradient is strong while the external voltage is relatively small, $v = 2$). Importantly, all the curves for various $n_L / n_R$ in Fig.6$(b)$ show the maximums in magnitude at the same value of $u_0$ that exactly coincides with the resonant potential $u_m$ determined in Fig.6$(a)$. Thus, we arrive at an important remark: at a given $q$ and $T$, the resonant potential is entirely determined by the channel structure parameters. It is an intrinsic property of the channels and can not be affected by the external factors such as the external voltage or the particle concentration gradient. 

\begin{figure} [t]
\includegraphics{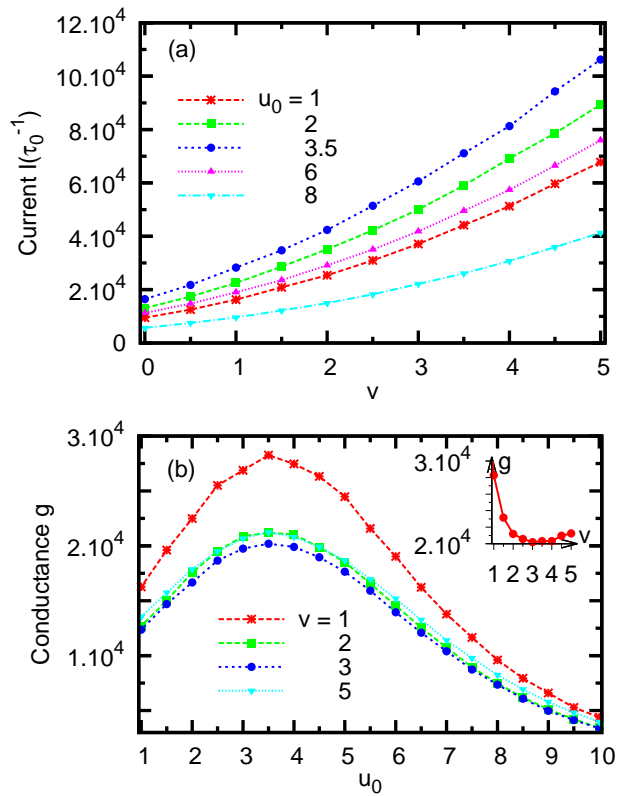}
\caption{
(color online) $(a)$ The current $I$ is plotted versus the voltage $v$ [I-V characteristics] for channels with different potentials $u_0$ (indicated in the figure). All I-V curves show a sublinear behavior. $(b)$ The channel ion conductance $g$ as a function of the channel potential $u_0$ at various voltages $v$ (given in the figure). At any $v$ the conductance always has the maximum at the same $u_0 = u_m \approx 3.5$. Inset: $g$ versus $v$ for the channel with  resonant potential $u_0 = u_m$.
}
\label{fig7ab}
\end{figure} 

Thus, in the present model, CFPM appears to be a self-optimized property of biological channels: to facilitate the transmembrane particle movement, the channels should be intrinsically optimized with appropriate structure parameters. Additionally, facilitating the transmembrane transport is very selective in the sense that a channel of definite structure parameters can facilitate the transmembrane transport of only particles of proper valence at corresponding temperatures.

Furthermore, we demonstrate in Fig.7$(a)$ the current-versus-voltage curves, $I(v)$ (I-V) - characteristics, extracted from Fig.5 for several channels of different $u_0$. The highest curve describes the I-V characteristics of the resonantly self-optimized channel with $u_0 = u_m$. It is clear that all the $I(v)$-curves presented in this figure are nonlinear, indicating the non-Ohmic property of the channel model studied. In this case, the channel ion conductance, defined as the ratio of $I$ to $v$ \cite{wilson}, becomes dependent on the applied voltage. The $I(v)$-curves in Fig.7$(a)$ reveal that as $v$ increases the conductances $g = I / v$ decrease fast first at small $v$, reach a minimum at $v \approx 2.5 - 3$, and then slightly increase at higher $v$ (see, for example, the inset in Fig.7$(b)$ for the case $u_0 = u_m$). The voltage, where the conductance gets minimal, depends on the potential $u_0$ and the reservoir particle concentrations. To look for a possible relation between the channel ion conductance and the resonant channel potential $u_m$ associated with the current [Fig.6], we present in Fig.7$(b)$ the conductances $g$ calculated for channels of different potentials $u_0$ at the same voltage $v$. Remarkably, at any $v$ the conductance $g$ always has the maximum at the same $u_0 = u_m$ as that identified in Fig.6 for the current. So, Fig.7$(b)$ gives one more demonstration for the resonantly self-organized property of channels in facilitating the transmembrane particle movement. The $v$-dependence of $g$ in this figure is related to the sublinear behavior of the I-V curves in Fig.7$(a)$ as just discussed above (see the Inset in Fig.7$(b)$). Note that such the sublinearity of calculated $I-V$ curves qualitatively resembles experimental data reported Refs.\cite{anders,busath}. 

\begin{figure} [t]
\includegraphics{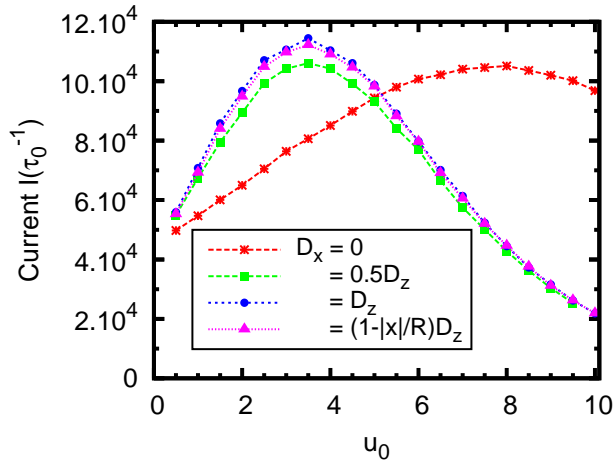}
\caption{
(color online) The $I(u_0 )$-curves for different $D_x$ (indicated in the figure) are compared to show the role of the transverse diffusion in the 2D-diffusion model considered [$v = 5$, $D_z = 0.5 D_0$, other parameters are the same as in Fig.2].
}
\label{fig8}
\end{figure} 

Finally, Fig.8 compares the $I(u_0 )$-curves obtained for different $D_x$ in showing the role of the transverse diffusion in the 2D-diffusion model under study. Four cases presented are: (1) $D_x = 0$, implying the 1D-diffusion, (2) $D_x = 0.5 D_z$, implying an anisotropically 2D-diffusion with $D_x$ constant and smaller than $D_z$, (3) $D_x = D_z$, implying an isotropically 2D-diffusion, and (4) $D_x = [1 - (|x| / R)] D_z$ used in this work (see corresponding symbols given in the figure). Obviously, the curve in the case of 1D-diffusion is largely separated from the rest, showing an essential role of the transverse diffusion. In this limiting case there has also the maximum in the $I(u_0 )$-curve, however, the current peak is lower and the resonant potential is much larger ($\approx 8$), compared to those for 2D-diffusion models. Interestingly, all the three 2D-diffusion $I(u_0 )$-curves with $D_x$ different on the behavior or the value show very similar forms with the same resonant potential $u_m = 3.5$. In addition, the isotropic 2D-diffusion model, $D_x = D_z$, provides the highest current peak. 
    
\section{Conclusions} 
We have considered an anisotropic 2D-diffusion of a charged molecule (particle) through a large biological channel under an external voltage. Connecting the two reservoirs with different particle concentrations, the channel is modeled as a rigid cylinder characterized by the three structure parameters: the radius, the length, and the surface density of the negative charges of channel interior-lining. These negative charges induce inside the channel a potential that is uniquely determined by the channel structure parameters and that critically affects the transmembrane particle movement. The suggested model is rather phenomenological so that the channel-induced potential can be calculated exactly. Nevertheless, it serves well to gain an understanding of the physical mechanism of the channel-facilitated particle movement. More detailed quantitative models are required to describe concrete realistic biological channels (see for example, \cite{treptow}). 

Our study is concentrated on showing the influences of this channel-induced potential and the external voltage on the typical dynamical characteristics of particles such as the translocation probabilities, the average translocation times, the net current, and the channel ion conductance. It was shown that while the external voltage does not cause any especial effect, the channel potential increases both the translocation probabilities and the average translocation times. And, surprisingly, studies demonstrated a single average translocation time that is equally applied for the particles passing the channel in two contrary directions, regardless of even the directed influence of the external voltage.

The most interesting result was appeared in examining the particle current. It was shown that at a given temperature the channel with appropriate structure parameters can induce the resonant potential that effectively facilitates the transmembrane movement of the particles of a given valence, resulting in a very large net particle current across the channel. In other words, to facilitate the transmembrane particle movement the channel should be naturally self-optimized so that its potential coincides with the resonant one. The resonant potential is an intrinsic characteristics of the channel and facilitating the transmembrane particle movement is an intrinsic property of biological channels, independent of the external factors such as the external voltage or the particle concentration gradient. In addition, the observed CFPM is very selective in the sense that a channel of definite structure parameters can facilitate the transmembrane movement of only particles of proper valence at corresponding temperatures. Calculated current-voltage characteristics also show that the channel model is non-Ohmic. The full characteristics of conductance exhibit an absolute maximum at the same resonant channel potential as that identified in the currents. 

It should be conclusively noted that all the results presented above are principally related to the considered single particle model, neglecting all the effects associated with the many-particle couplings, the particle size, and the potential induced by particle itself. So, these results might be served as an argument for further studies. \\ \\
{\bf Acknowledgments}. V.L.N. gratefully acknowledges a generous hospitality from Institute for Bio-Medical Physics in Hochiminh City, where this work has been done.    


\newpage
\begin{thebibliography}{99}
\bibitem{essent} B. Alberts, D. Bray, K. Hopkin, A. Johnson, J. Lewis, M. Raff, K. Roberts, P. Walter, {\sl Essential Cell Biology}, $4th$ edn. (Garland Science, Taylor \& Francis, New York - London, 2014).
\bibitem{hille} B. Hille, {\sl Ionic Channels of Excitable Membranes}, $3rd$ edn. (Sinauer Associates, Sunderland, 2001).
\bibitem{wickner} W. Wickner and R. Schekman, {\sl Science} {\bf 310}, 1452 (2005).
\bibitem{lerche} H. Lerche, K. Jurkat-Rott, and F. Lehmann-Horn, {\sl Am. J. Med. Genet.} {\bf 106}, 146 (2001).
\bibitem{marban} E. Marban, {\sl Nature} {\bf 415}, 213 (2002).
\bibitem{payan} J. Payandeh, T. Scheuer, N. Zheng, and W.A. Catterall, {\sl Nature} {\bf 475}, 353 (2011).
\bibitem{meller} A. Meller, {\sl J. Phys.: Condens. Matter} {\bf 15}, R581 (2003). 
\bibitem{moham} M.M. Mohammad, S. Prakash, A. Matouschek, L. Movileanu, {\sl J. Am. Chem. Soc.} {\bf 130}, 4081 (2008). 
\bibitem{neher} E. Neher and B. Sakmann, {\sl Nature} {\bf 260}, 799 (1976).
\bibitem{berezh1} A.M. Berezhkovskii, M.A Pustovoit, and S.M. Bezrukov, {\sl J. Chem. Phys.} {\bf 116}, 3943 (2003).
\bibitem{berezh2} A.M. Berezhkovskii, M.A Pustovoit, and S.M. Bezrukov, {\sl J. Chem. Phys.} {\bf 119}, 9952 (2002).
\bibitem{berezh3} A.M. Berezhkovskii, S.M. Bezrukov, {\sl Biophys. J.} {\bf 88}, L17 (2005).
\bibitem{berezh4} A.M. Berezhkovskii, S.M. Bezrukov, {\sl Chem. Phys.} {\bf 319}, 342 (2005).
\bibitem{bauer} W. R. Bauer and W. Nadler, {\sl PNAS} {\bf 103}, 11446 (2006)
\bibitem{kolomei1} A.B. Kolomeisky, {\sl Phys. Rev. Lett.} {\bf 98}, 048105 (2007).
\bibitem{kolomei2} A.B. Kolomeisky and K. Uppulury, {\sl J. Stat. Phys.} {\bf 142}, 1268 (2011).
\bibitem{bezruk} S. M. Bezrukov, A. M. Berezhkovskii, and A. Szabo, {\sl J. Chem. Phys.} {\bf 127}, 115101 (2007).
\bibitem{paglia} S. Pagliara, C. Schwall, and U.F. Keyser, {\sl Adv. Mater.} {\bf 25}, 844 (2013).
\bibitem{dettmer} S. L. Dettmer, S. Pagliara, K. Misiunas, and U. K. Keyser, {\sl Phys. Rev. E} {\bf 89}, 062305 (2014)
\bibitem{correa} A.M. Correa, F. Bezanilla, and R. Latorre, {\sl Biophys. J.} {\bf 61}, 1332 (1992).
\bibitem{erdem} R. Erdem and E. Aydiner, {\sl Phys. Rev. E} {\bf 79}, 031919 (2009).
\bibitem{anders} O.S. Anderson, R.E. Koeppell, {\sl Physiol. Rev.} {\bf 72}, S89 (1992).
\bibitem{busath} D.D. Busath, C.D. Thulin, R.W. Hendershot, L.R. Phillips, P. Maughan, C.D. Cole, N.C. Bingham, S. Morrison, L.C. Baird, R.J. Hendershot, M. Cotten, T.A. Cross, {\sl Biophys. J.} {\bf 75}, 2830 (1998).
\bibitem{graf} P. Graf, M.G. Kurnikova, R.D. Coalson, and A. Nitzan, {sl J. Phys. Chem. B} {\bf 108}, 2006 (2004).
\bibitem{corry} B. Corry and S.-H. Chung, {\sl Cell. Mol. Life Sci.} {\bf 63}, 301 (2006).
\bibitem{note02} Whereas the dielectric constant of the water in the channel interior is often assumed to be lower than the bulk value, generally, it is still remains an issue due to the lack of experimentally solid data \cite{graf}. Ignoring all these unsolved complications, for definition, we choose for $\epsilon$ the same value of the dielectric constant of the bulk water, $\epsilon \approx 80$. It is important to note that though the magnitude of the potential $U$ in eq.(3) does depend on $\epsilon$, the $U(x,z)$-behavior does not, and therefore, all qualitative conclusions of this work are unaffected by a choice of $\epsilon$-value.
\bibitem{moldyn} D. Frenkel and B. Smit, {\sl Understanding Molecular Dynamics Simulation}, $2^{nd}$ ed. ( Academic Press, San Diego - Tokyo, 2002).
\bibitem{euler} W.F. van Gunsteren and H.J.C. Berendsen, {\sl Molec. Phys.} {\bf 45}, 637 (1982).
\bibitem{fathi} M. Fathi and G. Stoltz, {\sl arXiv}: 1505.04905v1
\bibitem{maffeo} C. Maffeo, S. Bhattacharya, J. Yoo, D. Wells, and A. Aksimentiev, {\sl Chem. Rev.} {\bf 112}, 6250 (2012).
\bibitem{wilson} M.A. Wilson, T.H. Nguyen, and A. Pohorille, {\sl J. Chem. Phys.} {\bf 141}, 22D519 (2014).
\bibitem{treptow} W. Treptow and M. Tarek, {\sl Biophys. J.} {\bf 91}, L26 (2006); {\bf 91}, L81 (2006).
\end{thebibliography}
\end{document}